\documentclass[acmsmall,screen,nonacm]{acmart}

\AtBeginDocument{%
  }

\setcopyright{acmlicensed}
\copyrightyear{2018}
\acmYear{2018}
\acmDOI{XXXXXXX.XXXXXXX}
\acmConference[Conference acronym 'XX]{Make sure to enter the correct
  conference title from your rights confirmation email}{June 03--05,
  2018}{Woodstock, NY}
\acmISBN{978-1-4503-XXXX-X/2018/06}

\usepackage[utf8]{inputenc} 
\usepackage[T1]{fontenc}    
\usepackage{hyperref}       
\usepackage{url}            
\usepackage{booktabs}       
\usepackage{amsfonts}       
\usepackage{nicefrac}       
\usepackage{microtype}      
\usepackage{amsmath}
\usepackage{appendix}
\usepackage{graphicx}
\graphicspath{./figures/}
\usepackage{multirow}

\usepackage{float}
\floatstyle{plain}
\newfloat{listing}{htbp}{lop}
\floatname{listing}{Listing}

\usepackage{tabularx}
\usepackage{caption}
\usepackage{pgfplotstable}
\pgfplotsset{compat=1.17}
\usepackage{adjustbox}
\usepackage{arydshln} 

\usepackage[table,dvipsnames]{xcolor}
\usepackage{colortbl}

\usepackage{tcolorbox}
\usepackage{listings}
\usepackage{pifont}  
\newcommand{\cmark}{\textcolor{green!60!black}{\ding{51}}} 
\newcommand{\xmark}{\textcolor{red!70!black}{\ding{55}}} 

\usepackage{marginnote}

\newcommand{\commentout}[1]{}

\author{Daniel Koh}
\affiliation{%
  \institution{Mohamed bin Zayed University of Artificial Intelligence}
  \country{United Arab Emirates}
}

\author{Yannic Noller}
\affiliation{%
  \institution{Ruhr-Universität Bochum}
  \country{Germany}
}

\author{Corina S. Pasareanu}
\affiliation{%
  \institution{Carnegie Mellon University}
  \country{United States of America}
}

\author{Adrians Skapars}
\affiliation{%
  \institution{University of Manchester}
  \country{United Kingdom}
}

\author{Youcheng Sun}
\affiliation{%
  \institution{Mohamed bin Zayed University of Artificial Intelligence}
  \country{United Arab Emirates}
}

\begin{document}

\title{Worst-Case Symbolic Constraints Analysis and Generalisation with Large Language Models}

\renewcommand{\shortauthors}{Koh et al.}

\begin{abstract}

Large language models (LLMs) have demonstrated strong performance on coding tasks such as generation, completion and repair, but their ability to handle complex symbolic reasoning over code still remains underexplored. We introduce the task of worst-case symbolic constraints analysis, which requires inferring the symbolic constraints that characterise worst-case program executions; these constraints can be solved to obtain inputs that expose performance bottlenecks or denial-of-service vulnerabilities in software systems. We show that even state-of-the-art LLMs (e.g., GPT-5) struggle when applied directly on this task. To address this challenge, we propose \textbf{WARP}, an innovative neurosymbolic approach that computes worst-case constraints on smaller concrete input sizes using existing program analysis tools, and then leverages LLMs to generalise these constraints to larger input sizes.

Concretely, WARP comprises: (1) an incremental strategy for LLM-based worst-case reasoning, (2) a solver-aligned neurosymbolic framework that integrates reinforcement learning with SMT (Satisfiability Modulo Theories) solving, and (3) a curated dataset of symbolic constraints. Experimental results show that WARP consistently improves performance on worst-case constraint reasoning. Leveraging the curated constraint dataset, we use reinforcement learning to fine-tune a model, \texttt{WARP-1.0-3B}, which significantly outperforms size-matched and even larger baselines. These results demonstrate that incremental constraint reasoning enhances LLMs’ ability to handle symbolic reasoning and highlight the potential for deeper integration between neural learning and formal methods in rigorous program analysis.
\end{abstract}

\ccsdesc[500]{Computing methodologies~Artificial intelligence}
\ccsdesc[300]{Theory of computation~Automated reasoning}
\ccsdesc[200]{Software and its engineering~Constraint and logic programming}
\ccsdesc[100]{Theory of computation~Verification by model checking}

\keywords{large language models, symbolic reasoning, worst-case constraint reasoning, SMT solving, reinforcement learning, program analysis, neurosymbolic methods}


\maketitle

\begin{figure}[t]
    \definecolor{flowchart_custom_color_green}{HTML}{E7F9E5}
    \centering
    \includegraphics[width=\textwidth]{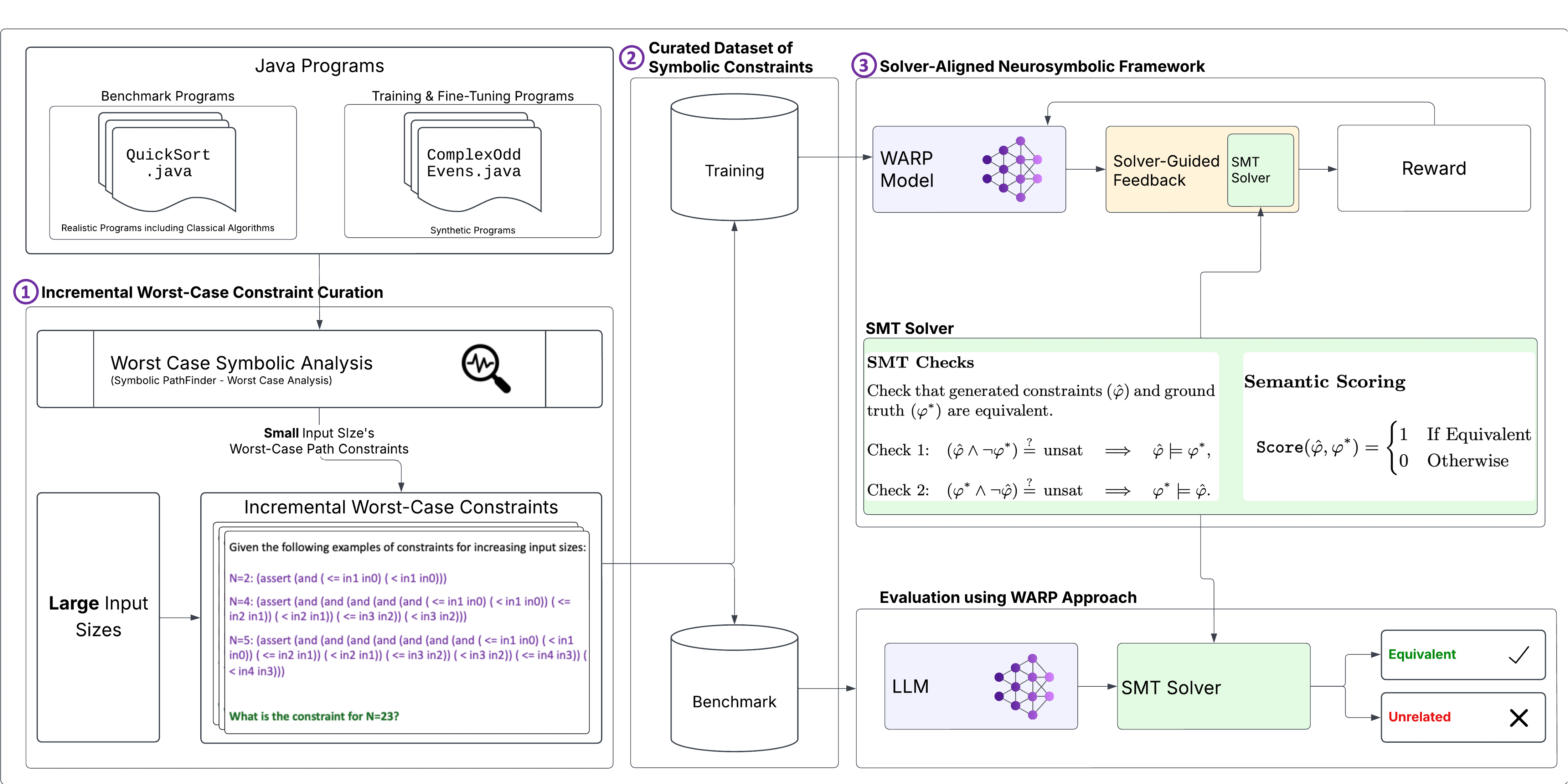}
    \caption{Overview of \texttt{WARP} for worst-case execution analysis. 
    \textbf{1. Incremental Worst-Case Path Constraints Curation}: Java programs are analysed with \texttt{SPF-WCA} to generate worst-case path constraints for smaller concrete input sizes, which serve as tractable constraints for extrapolating to larger input sizes. 
    \textbf{2. Curated Dataset and Benchmark of Symbolic Constraints}: Consolidates these constraints into a resource that both trains models on worst-case execution patterns and serves as a benchmark for evaluating generalisation beyond symbolic execution’s limits. 
    \textbf{3. Solver-Aligned Neurosymbolic Framework}: The WARP (policy) model generates constraint candidates from prompts, refined via solver-guided feedback and validated with an \texttt{\colorbox{flowchart_custom_color_green}{SMT Solver}} for semantic alignment.}

    \label{fig:flowchart}
\end{figure}

\section{Introduction}
\label{sec:introduction}
Understanding the worst-case algorithmic complexity of software systems has many practical applications, ranging from compiler optimisations, finding and fixing performance bottlenecks, or improving cybersecurity in the presence of denial-of-service attacks. Profilers are typically used to find performance bottlenecks; however, they are inherently incomplete, as they are limited by the number of test inputs used. Symbolic execution techniques, such as WISE \citep{wise-benchmarks} and SPF-WCA \citep{spf-wca}, attempt exhaustive analysis but face severe scalability limitations, due to the exponential growth of possible execution paths.  
%
%
Meanwhile, large language models (LLMs), exemplified by GPT-5 \citep{openai-gpt5-systemcard}, have demonstrated remarkable capabilities in a variety of programming tasks, including code generation, completion, and repair \citep{poldrack2023ai,fu2023chatgpt}. However, their sub-symbolic architecture--based on statistical pattern recognition rather than explicit logical reasoning--is not immediately suited for the kinds of rigorous analysis required in worst-case complexity reasoning or formal software verification.
This gap has catalysed the field of neurosymbolic AI, which reconciles the flexibility of neural inference with the precision of symbolic reasoning. Recent work in the automated theorem proving domain has applied various approaches like reinforcement learning, retrieval-augmented generation and integration of external symbolic tools (often called ``hammers'') to uplift the capabilities of LLMs in the neurosymbolic reasoning dimension \citep{xin2024deepseekproverv15harnessingproofassistant,yang2023leandojotheoremprovingretrievalaugmented,jiang2022thorwieldinghammersintegrate}.

Building on this momentum, we introduce {\em \textbf{W}orst-case \textbf{A}symptotic \textbf{R}easoner for \textbf{P}ath constraints} (WARP), a unified neurosymbolic framework for the synthesis and verification of worst-case path constraints, as illustrated in Figure \ref{fig:flowchart}. WARP combines symbolic program analysis, reinforcement learning with solver-guided feedback, and rigorous benchmarking to improve the asymptotic reasoning capabilities of LLMs.
%
%
%
Our work advances neurosymbolic reasoning for program analysis by addressing the challenge of integrating formal verification with LLMs, and it makes the following contributions:

\paragraph{Incremental Worst-Case Constraint Reasoning}
As one of our primary contributions, we introduce an \emph{incremental} inference strategy for worst-case constraint reasoning with LLMs: obtain solver-checked worst-case constraints for \emph{small} input sizes using existing analysis tools, then (train and) prompt LLMs to \emph{generalise} these constraint patterns to \emph{larger} inputs. This reframes the task as semantics-preserving pattern induction, sharply reduces the effective search space, mitigates path explosion, and yields substantial gains over direct prompting on large inputs.

\paragraph{Neurosymbolic Approach for Worst-Case Complexity Analysis} At the core of the WARP framework is a novel learning paradigm, which we term \textit{Reinforcement Learning from Solver-Guided Feedback} (RLSGF). This method fine-tunes LLMs to generate symbolic constraints by leveraging feedback from SMT solvers, for the synthesis of semantically valid and generalisable SMT formulas for analysing software programs. Rather than exhaustively exploring an exponentially growing search space, WARP features training on tractable, small-scale program input constraints, enabling the model to extrapolate constraint patterns to larger inputs—where traditional tools often fail due to path explosion.

\paragraph{Fine-Tuning and Benchmarking Datasets}
To advance and evaluate worst-case–constraint synthesis, we introduce two complementary resources. First, a \emph{fine-tuning dataset} of synthetically generated programs and their worst-case path constraints is designed to elevate the symbolic-reasoning skills of LLMs. Secondly, a first-of-its-kind benchmark that assesses the semantic fidelity of generated constraints, formal correctness, and generalisation. By focusing on the under-explored task of LLM-driven synthesis of worst-case constraints, the benchmark provides a rigorous, reproducible testbed for future work in neurosymbolic reasoning for WCA spectrum.


\paragraph{A Trained Model for Worst-Case Symbolic Constraint Reasoning}
    We develop and also release \texttt{WARP-1.0-3B}, an LLM fine-tuned using the framework to produce symbolic constraints that generalise across input sizes and outperform baselines on asymptotic prediction tasks.

\section{Related Work}

\subsection{Symbolic Program Analysis}
Early program model checkers like Java PathFinder (JPF) \citep{jpfhavelund} systematically explore program execution states to find bugs, but face scalability issues in complex programs. Symbolic PathFinder (SPF) \citep{10.1145/1390630.1390635} extended JPF by executing programs on symbolic inputs instead of concrete values, using constraint solvers to solve path conditions. This enabled automated test generation and analysis of Java bytecode with high coverage. Symbolic execution inherently suffers from path explosion in which the number of feasible paths grows exponentially with the number of branches. Worst-case analysis (WCA) further accentuates this problem, as it often requires the exploration of an asymptotically exponential number of execution paths, thereby rendering exhaustive analysis computationally intractable. Prior works introduced heuristics in attempt to mitigate path explosion. For example, \emph{WISE} \citep{wise-benchmarks} uses symbolic execution at small input scales and learns simple “generator” policies (e.g. always-true branches) to steer exploration toward worst-case paths at larger scales. Meanwhile, SPF-WCA \citep{spf-wca} built as an extension to the SPF tool supports more complex branching strategies (tracking branch history) to ensure worst-case paths are explored. While these approaches prune the search, they remain limited by the exponential path growth and heavy reliance on solvers.

\subsection{LLMs for Program Analysis}
The emergence of LLMs pretrained on code (e.g. GPT-4, Codex) has opened new avenues for program reasoning. Modern LLMs can generate syntactically correct code and even perform simple reasoning about program behaviour in-context. This has spurred research into using LLMs for software verification and analysis tasks beyond code synthesis. For instance, \cite{pmlr-v202-pei23a} investigated whether LLMs can infer loop invariants and other program properties important for verification. Their study demonstrated that a sufficiently powerful LLM can propose relevant invariants for synthetic Java programs, hinting at the model’s reasoning ability. Similarly, \cite{tihanyi2024newerasoftwaresecurity} showed that LLMs can assist in vulnerability repair by generating candidate patches which are then validated with formal verification tools. Despite these promising results, purely non-symbolic approaches do not offer formal correctness guarantees. That is, an LLM’s output (e.g., an invariant or a code patch) might seem plausible while being erroneous. Consequently, because sub-symbolic LLMs operate as probabilistic next-token predictors \citep{brown2020languagemodelsfewshotlearners}, it is necessary to deploy an external checker (e.g., an SMT solver or verifier) to validate these outputs.

\subsection{Neurosymbolic Reasoning}
Increasingly, researchers are exploring neurosymbolic methods that integrate neural networks with symbolic reasoning to capitalise on the strengths of each. In the context of program verification, neurosymbolic systems use learning to guide or supplement traditional formal analyses.  GPT-f \citep{polu2020generativelanguagemodelingautomated} was an early landmark, integrating a transformer-based language model into the Metamath \citep{metamath} formal system as a proof search agent. GPT-f demonstrated that training using the prover’s success to update a value function guiding search can continuously improve theorem-proving performance. Subsequent systems have expanded on this idea. For example, \cite{zombori2021findinglongerproofs} applied deep reinforcement learning to guide a theorem prover, employing graph neural networks to represent logical formulae, with the aim of handling long, complex proofs. Techniques inspired by AlphaZero’s self-play have also been adapted to propose proof steps and receive verification feedback at each step. While effective, these RL approaches can be computationally expensive due to the need to attempt many proof paths and compute rewards from successful proofs. 

An alternative direction has been to use pre-trained LLMs with post-training methods. One notable example is LeanDojo \citep{yang2023leandojotheoremprovingretrievalaugmented}, an open-source playground that enables LLMs to interact with the Lean theorem prover \citep{lean_theorem_proving}. LeanDojo provides an interface and dataset of proofs, allowing an LLM-based agent to propose proof steps and query the proof assistant. By augmenting an LLM with a retrieval mechanism over a Math library, LeanDojo can select relevant premises and tactics, significantly improving success rates in formal proofs compared to untutored LLM reasoning. DeepSeek Prover \citep{xin2024deepseekproverv15harnessingproofassistant} pushes this synergy further by incorporating reinforcement learning signals from a proof assistant during training. DeepSeek uses Lean to provide a reward signal to fine-tune an LLM for theorem proving. In doing so, the neural model learns to align its generation with formal proof requirements, blending learned intuition with strict verification.
Such efforts highlight that combining LLMs with symbolic evaluators (proof assistants, static analysers, solvers) can yield systems that are both creative and more reliable. Our work falls into this neurosymbolic spectrum, where we study and evaluate the application of LLM to synthesise worst-case complexity path constraints in \texttt{SMT-LIB v2} (SMT)~\citep{BarFT-SMTLIB} format. We further formally verify that the synthesised constraints are semantically equivalent to the worst-case path constraints using the SMT solver \texttt{Z3}~\citep{z3-demoura}. The verification is used to guide training and evaluate model capability.
\section{Problem Formulation}
\label{sec:problem formulation}
Symbolic worst-case analysis identifies input conditions under which software exhibits its maximal resource usage -- whether time, memory, or instruction count. Identifying these conditions is crucial to prevent vulnerabilities and bottlenecks in real-world applications. However, existing symbolic execution tools, such as SPF-WCA, face significant scalability limitations due to the exponential path growth.
Symbolic execution explores the program paths by solving the path constraints generated from conditional branches. The worst-case complexity emerges for input size \(n\) from the paths \(\Pi(n)\) yielding the maximal resource consumption\(\underset{\pi\in\Pi(n)}{\max} C(\pi)\) where \(C(\pi)\) is the cost function for path \(\pi\).

\lstdefinestyle{javaStyle}{
  language=Java,
  numbers=left,
  numberstyle=\tiny\color{gray},
  numbersep=10pt,
  xleftmargin=2em,
  frame=none,
  rulecolor=\color{black},
  basicstyle=\ttfamily\small,
  keywordstyle=\color{violet}\bfseries,
  commentstyle=\color{gray}\itshape,
  stringstyle=\color{orange},
  showstringspaces=false,
  tabsize=2,
  breaklines=true,
  backgroundcolor=\color{white}
}
\begin{listing}[t]
\begin{lstlisting}[
  style=javaStyle,
  firstnumber=1,
  caption={Excerpt of the \texttt{QuickSort} function taken from the benchmark, which implements a divide-and-conquer approach to sort an array by partitioning it around a pivot element and recursively sorting the resulting subarrays. This method typically runs in \(O(n\;log(n))\) time on average, but may degrade to \(O(n^2)\) in the worst case when the pivot selection produces highly unbalanced partitions.},
  label={lst:quicksort}
]
public class QuickSort {
    public static void quickSort(int[] arr, int low, int high) {
        if (low < high) {
            int pivotIndex = partition(arr, low, high);
            quickSort(arr, low, pivotIndex - 1);
            quickSort(arr, pivotIndex + 1, high);
        }
    }
    
    public static int partition(int[] arr, int low, int high) {
        int pivot = arr[high];
        int i = low - 1;
        for (int j = low; j < high; j++) {
            if (arr[j] <= pivot) {
                i++;
                int temp = arr[i];
                arr[i] = arr[j];
                arr[j] = temp;
            }
        }
        int temp = arr[i+1];
        arr[i+1] = arr[high];
        arr[high] = temp;
        return i + 1;
    }
}
\end{lstlisting}
\end{listing}

\subsection{Motivating Example}
Consider the example \texttt{QuickSort} (see \autoref{lst:quicksort}). Worst-case analysis can derive constraints for small inputs (e.g. \(n=3\)) succinctly represented as SMT formula: 
\[
\texttt{(assert (and  (<= in0 in2)  (<= in1 in2)  (<= in0 in1)))}
\]

As input size grows (e.g. \(n=12\)), the symbolic execution tool's computation becomes increasingly intensive, quickly exceeding thousands of logical clauses. In particular, this represents a \(2^{12}=4096\)-fold increase in the number of execution paths compared to input \(n=3\) above. 

Even with optimised symbolic methods, like history-preserving guidance, this quickly becomes computationally infeasible at larger input sizes, requiring paradoxical amounts of computational resources and time. Analysing \texttt{QuickSort} at \(n=30\) entails exploring \(2^{30}\approx10^9\) paths, translating to impractical CPU time and memory usage, rendering exhaustive methods computationally infeasible.

\begin{table}
  \centering
  \caption{Results of Vanilla Prompting with GPT-5}
  \label{tab: vanilla prompting results}
  \setlength{\tabcolsep}{4pt}
  \renewcommand{\arraystretch}{1.05}

  \begin{tabular*}{\linewidth}{@{\extracolsep{\fill}} p{0.38\linewidth} *{5}{c}}
    \toprule
    \multirow{2}{*}{\textbf{WARP Benchmark Programs}} & \multicolumn{5}{c}{\textbf{Input Size}} \\
     & \textbf{2} & \textbf{4} & \textbf{8} & \textbf{16} & \textbf{30} \\
    \midrule
    ArrayTwister               & \xmark & \xmark & \xmark & \xmark & \xmark \\
    BinarySearch               & \xmark & \cmark & \cmark & \cmark & \xmark \\
    BinarySearchTreeHeight     & \cmark & \xmark & \cmark & \cmark & \cmark \\
    BinaryTreeSearch           & \xmark & \xmark & \xmark & \xmark & \xmark \\
    BubbleSort                 & \cmark & \cmark & \cmark & \xmark & \xmark \\
    CaseFlipper                & \xmark & \xmark & \xmark & \xmark & \xmark \\
    Collatz                    & \xmark & \xmark & \xmark & \xmark & \xmark \\
    ComplexStateMachineParser  & \xmark & \xmark & \xmark & \xmark & \xmark \\
    Dijkstra                   & \xmark & \xmark & \xmark & \xmark & \xmark \\
    DizzyRamp                  & \xmark & \xmark & \xmark & \xmark & \xmark \\
    GreedyStepper              & \xmark & \xmark & \xmark & \xmark & \xmark \\
    KnapsackSolver             & \xmark & \xmark & \xmark & \xmark & \xmark \\
    MazeSolver                 & \xmark & \xmark & \xmark & \xmark & \xmark \\
    MergeSort                  & \xmark & \xmark & \xmark & \xmark & \xmark \\
    NaiveFibonacci             & \xmark & \xmark & \xmark & \xmark & \xmark \\
    QuickSort                  & \cmark & \xmark & \xmark & \xmark & \xmark \\
    RampUp                     & \xmark & \xmark & \xmark & \xmark & \xmark \\
    SortedListInsert           & \xmark & \xmark & \xmark & \xmark & \xmark \\
    SubarraySumFinder          & \cmark & \xmark & \xmark & \cmark & \xmark \\
    TowerOfHanoi               & \xmark & \xmark & \xmark & \xmark & \xmark \\
    \bottomrule
  \end{tabular*}

  \vspace{0.3em}
  {\footnotesize Notes: \cmark = TRUE, \xmark = FALSE}
\end{table}


\lstset{
  basicstyle=\ttfamily\footnotesize,
  escapeinside={(*@}{@*)}, 
  breaklines=true,
  frame=lines
}

\begin{lstlisting}[float,caption={Vanilla prompt structure for worst-case constraint reasoning. 
\textcolor{red}{[\texttt{WARP benchmark Program}]} denotes the placeholder for a single WARP benchmark program (see \autoref{subsec:constraint-curation}). 
\textcolor{red}{[\texttt{Input Size}]} is an integer in the set \(n=\{2,4,8,16,30\}\).},
label={lst:vanilla-prompts}]
(*@\textcolor{red}{[WARP Benchmark Program]}@*)

What is the worst-case time complexity of the Java program above. 
Also, provide us the corresponding constraint in SMT2 format 
for input size n:(*@\textcolor{red}{[Input Size]}@*)

Give the final SMT2 constraint at the end starting with 'Answer:'
\end{lstlisting}

\subsection{Vanilla Prompting}
\label{subsec:RQ1}


We first explore the direct application of an LLM to generate SMT worst-case constraints for the Java programs.
The LLM is given the program and it is instructed to output a syntactically correct, canonical SMT assertion (see \autoref{lst:vanilla-prompts}).
%
To quantify performance under \textit{vanilla prompting}, we evaluated a state-of-the-art proprietary LLM, \texttt{GPT-5} \citep{openai-gpt5-systemcard}, across all $20$ programs in the \textsc{WARP} benchmark (see \autoref{subsec:constraint-curation}). For each program, we sampled five input sizes \(\{2,4,8,16,30\}\) and tasked the model with generating the worst-case path constraints at each size. A generated constraint is counted as correct only if it is both: (1) syntactically valid and (2) semantically equivalent to the ground-truth worst-case constraint verified by the Z3 solver. This dual criterion ensures that correctness reflects true logical equivalence rather than superficial string similarity.

As summarised in \autoref{tab: vanilla prompting results}, \texttt{GPT-5} achieves only \(13/100\) in overall accuracy. Most outputs are syntactically well-formed but fail semantic equivalence checks, underscoring the difficulty of worst-case constraint reasoning: without additional structure, even current state-of-the-art models cannot reliably generate correct constraints.
These findings motivate the incremental neurosymbolic reasoning strategy of WARP: rather than requiring the model to induce full constraints from scratch, we scaffold learning with solver-verified instances at smaller input sizes, enabling systematic extrapolation to larger and otherwise infeasible regimes.

\subsection{Augmenting Symbolic Execution with LLMs}
We propose a neurosymbolic approach that uses SPF-WCA and LLMs in an incremental fashion. 
Concretely, we use SPF-WCA to obtain constraints at small input sizes and we task the LLM to:
\begin{enumerate}
    \item Learn constraint patterns from those tractable input sizes.
    \item Extrapolate these patterns to predict constraints for significantly larger inputs.
\end{enumerate}

\begin{definition}[Incremental Worst-Case Constraint Reasoning]
\label{def:incr}
Given a sequence of solver-derived constraint sets 
$\{\Phi_{n_1}, \Phi_{n_2}, \dots, \Phi_{n_k}\}$ 
for small input sizes $n_1 < n_2 < \cdots < n_k$, infer a constraint set $\hat{\Phi}_n$ for a larger input size $n \gg n_k$ such that $\hat{\Phi}_n \equiv \Phi_n$ with respect to solver-based semantic equivalence.
\end{definition}

This incremental formulation reframes worst-case analysis as a \emph{pattern induction problem}: instead of exploring all paths for large $n$, the learner generalises from small-scale constraints to approximate or reproduce $\Phi_n$. This reduces the exponential search space, mitigates path explosion, and enables analysis at input scales where symbolic execution alone is infeasible.

\section{Methodology}
\label{sec:methodology}

Figure \ref{fig:flowchart} gives a high-level overview of WARP. Note that WARP includes a benchmark designed to evaluate constraint generalisation, and it is applied to finetune an LLM through reinforcement learning to synthesise SMT constraints for our neurosymbolic task. Our results show that the incremental reasoning approach significantly improves LLMs’ ability to solve worst-case constraint reasoning tasks. The fine-tuned \texttt{WARP-1.0-3B} model also matches the performance of much larger proprietary and open-source baselines in terms of both accuracy and semantic validity of generated constraints. Therefore, our results underscore WARP's effectiveness as a lightweight neurosymbolic alternative. 

\subsection{Incremental Worst-Case Path Constraints Curation}
\label{subsec:constraint-curation}

The defining feature of WARP is its \emph{incremental} approach to worst-case constraint reasoning. 
Instead of asking an LLM to derive large-$n$ constraints from scratch, we first curate solver-verified constraints at small input sizes where symbolic execution is efficient and reliable. 
These small-$n$ constraints act as building blocks: they capture the recurring structural patterns of worst-case program behaviour and provide the context needed for the LLM to extrapolate to larger, harder cases.
Concretely, we use SPF-WCA~\citep{spf-wca} to analyse programs at tractable input sizes in its default configuration. SPF-WCA's default cost model utilises the depth of the symbolic execution path, i.e., the more branches with symbolic conditions are visited during the execution, the more costly such a path is.
For each program and size $n$, SPF-WCA yields worst-case path constraints in SMT format. 
By collecting these constraints across multiple small-$n$ values, we build sequences that illustrate how the constraint structure evolves as $n$ increases.

WARP leverages these curated constraints in an incremental fashion during both inference and training. 
During inference, curated examples at smaller $n$ values are included in the prompt, and the model is tasked with generating the constraint for a larger, held-out input size. 
During fine-tuning, they form the foundation of a dataset that familiarises LLMs with the syntax and semantics of SMT constraints while teaching them the progression of constraint patterns across input sizes. 
This setup explicitly isolates the ability to generalise incrementally, ensuring that observed performance gains reflect learned symbolic reasoning patterns rather than access to ground-truth large-$n$ solutions.

\subsection{Dataset Creation and Benchmarking}
\label{subsec:dataset}

In order to accommodate the novel application of LLMs for the prediction of worst-case path constraints, a corpus of Java programs and the worst-case execution path constraints that follow have been collected and formatted to enable the development of generalisation and predictive capabilities in LLMs through reinforcement learning.
For fine-tuning, we assembled a diverse dataset of 23,600 instances—comprising 17,700 training examples and 5,900 test examples—that span a broad range of symbolic constraints, from simple arithmetic checks to complex logical compositions. This collection includes both synthetic programs and examples derived from real-world applications. Notably, a portion of the dataset comes from SPF-WCA and the Badger tool's examples \citep{Noller_2018}, which capture a variety of behaviours representative of real-world applications.

\begin{figure}
    \centering
    \includegraphics[width=\linewidth]{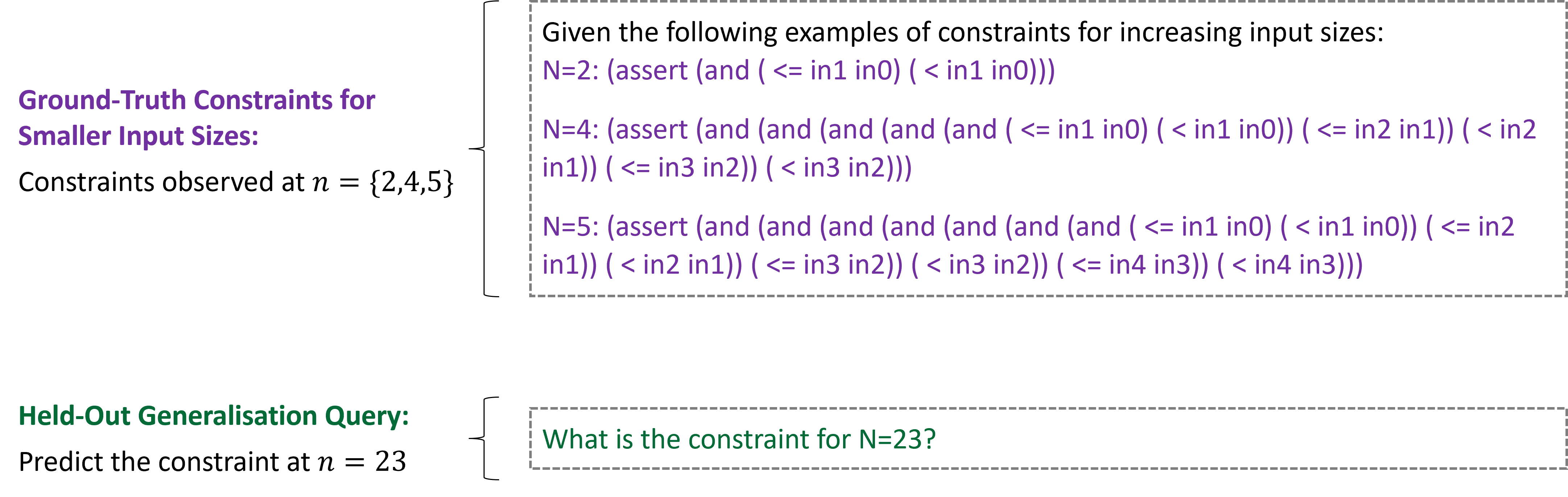}
    \caption{Structure of a single instance in WARP-benchmark.}
    \label{fig:warp-benchmark instance}
\end{figure}

Independent from our fine-tuning dataset, we compiled another test set consisting of over 600 instances from 20 Java programs. Each instance includes ground-truth SMT constraints for several smaller \(n\) values from a single Java program, designating a larger \(n\) as the held-out target the model must synthesise (see \autoref{fig:warp-benchmark instance}). By comparing the generated constraints against the held-out ground-truth, we directly measure a model’s ability to extrapolate the underlying symbolic rule rather than merely memorising observed patterns. These programs implement different algorithms, spanning array manipulation, searching, sorting, graph traversal, optimisation and recursion.

\paragraph{Defining Difficulty via Jumps in \(n\)}
We posit that \emph{difficulty} depends not only on the upper bound of \(n\) but also on the magnitude of the leap from the largest available candidate example to the target. For instance, employing examples at \(n=1,2,3\) and querying \(n=30\) can be considered more challenging than progressing through intermediate values such as (\(n=1,3,6,8,15\)) to \(n=30\). Larger leaps demand more extensive extrapolation, while smaller, incremental steps emphasise granular pattern replication.

\paragraph{Tiered Complexity and Model Limitations}
To accommodate both modest and extensive leaps without incurring path explosion, we classify each instance into small, medium, and large \emph{tiers}, each associated with distinct ranges of \emph{jump size}. As input size \(n\) increases, synthesising the corresponding worst-case constraints requires more reasoning and results in longer outputs, often nearing or exceeding the model’s maximum generation capacity. While providing more examples in the prompt can improve output quality by offering richer context, it also encourages the model to generate longer reasoning chains, which consume a greater portion of the output budget. This introduces a critical trade-off: richer prompts can improve intermediate reasoning, but increase the risk that the model truncates before fully emitting the final constraint. To mitigate this, we impose an upper limit on prompt size to ensure adequate capacity is preserved for generation. When the full set of example constraints exceeds this limit, we reduce the prompt to three representative examples—the first, the median, and the last. If the reduced prompt still exceeds the threshold, we exclude the instance entirely. Although this may conservatively filter out potentially solvable cases, it serves as a practical safeguard: if three examples already exceed the output limit, the constraint for a larger \(n\) is likely even longer and prone to truncation.

Note that this constraint is not only a technical safeguard but also a deliberate design choice intended to foster broader accessibility. By bounding prompt and output sizes, we make the benchmark approachable for smaller models, enabling research on efficient reasoning under limited resources. Looking forward, we view this as a foundation for extensible benchmarks that systematically explore trade-offs between model scale, prompt complexity, and output fidelity—paving the way for evaluating both scaling trends and compact, capable alternatives.

\paragraph{Final Benchmark Set and Scope}
After refining the benchmarking dataset through these constraints, we obtain a consolidated set of 671 instances. Each instance includes a prompt (featuring constraints at multiple smaller \(n\)) and asks for a constraint at a larger \(n\), alongside the corresponding solution. Restricting \(n\) to at most 30 allows for practical verification using the Z3 SMT solver \citep{z3-demoura}, ensuring each instance can be definitively classified as correct or incorrect. 

\subsection{Reinforcement Learning from Solver-Guided Feedback}
\label{sec:rl-enhanced model}

\begin{lstlisting}[float, caption={Template used for training WARP-1.0. (\textcolor{red}{\texttt{EXAMPLES}}) denotes the placeholder for specific constraint instances corresponding to smaller input sizes (e.g.  \texttt{N=1:(assert (<= in0 in1))}). (\textcolor{red}{\texttt{QUESTION}}) is replaced with the target input size for which the model is asked to generate the corresponding constraint (e.g., \texttt{30}).}, label={tab:RL Template}]
A conversation between User and Assistant.
The user asks a question, and the Assistant solves it.

User: Your role is to take a known pattern of symbolic constraints that represent the 
longest execution path of a program and generalize it for any given input size N. 
When you receive an input value N, you must generate a canonical SMT-LIB constraint 
string that adheres to the following rules:
(assert (op (op (op var_1 var_2)) (op (op var_3 var_4)) (op (op var_5 var_6)) (op var_7 var_8)))
where op is a logical operator (e.g., 'and', 'or', 'not') and var_i are variables or constants.
All per-variable constraints must be combined using a top-level (assert (and ...)) 
clause. The output must be in exact, canonical SMT-LIB format without extra 
commentary in the constraint string.
Show your work in <think> </think> tags. And return the final SMT-LIB constraint string in <answer> </answer> tags.
For example: <answer>(assert (and  ( >=  in0 97)  ( <=  in0 122)))</answer>.
Here are the known constraints:
(*@\textcolor{red}{[EXAMPLES]}@*)
What is the constraint for N=(*@\textcolor{red}{[QUESTION]}@*)?

Assistant: Let me solve this step by step.<think>
\end{lstlisting}

We also contribute \texttt{WARP-1.0-3B}, a \textbf{3B parameter model} fine-tuned to generate SMT constraints for worst-case execution. For fine-tuning, we adapt the computationally-efficient framework proposed by \citet{sheng2024hybridflow} with the following key components:

\paragraph{Training Template} 
We build upon \texttt{Qwen2.5-3B} \citep{qwen2025qwen25technicalreport} and adapt it using a task-specific prompt designed to elicit SMT constraint generation. As illustrated in 
\autoref{tab:RL Template}, the model is required to first reason through the problem before providing its final answer. This design choice, inspired by \citet{deepseekai2025deepseekr1incentivizingreasoningcapability}, avoids imposing content-specific biases and allows the model’s intrinsic problem-solving capabilities to emerge naturally.

\paragraph{Reward Modelling} 
We adopt a rule-based reward system that consists of two types of reward:
\begin{itemize}
    \item \textbf{Semantic Reward.}  
    For each predicted SMT constraint $\hat{\varphi}$ we call the \textsc{Z3} solver to check logical equivalence with the ground‑truth constraint $\varphi^{\star}$. If \textsc{Z3} proves $\hat{\varphi}\Leftrightarrow\varphi^{\star}$, the episode receives a positive reward; otherwise it receives no reward. This binary signal is cheap to compute and correlates well with downstream verification performance described in \autoref{subsec:evaluation and performance metrics}.
    \item \textbf{Syntactic Reward.}  
    A smaller bonus is granted when the output exactly matches the required template \texttt{\small <think>\dots</think><answer>\dots</answer>}.  
    Any deviation yields zero bonus, discouraging malformed outputs without allowing the model to trade correctness for formatting.
\end{itemize}
The overall reward \(R\) is a weighted sum of the two components:
\begin{equation}    
R = 0.1 \times \texttt{Syntactic\_Reward} + 0.9 \times \texttt{Semantic\_Reward}
\end{equation}

\noindent This reflects our prioritisation of logical correctness over surface-level fluency. We intentionally leave the content of the \texttt{<think>} block unscored, focusing evaluation budget on verifiable constraint generalisation rather than natural language reasoning quality.

\lstset{
  basicstyle=\ttfamily\footnotesize, 
  breaklines=true,
  breakatwhitespace=true,
  postbreak=\mbox{},
  breakindent=0pt,
  frame=lines,
  escapeinside={(*@}{@*)},
}



\paragraph{Reinforcement Learning Algorithm} 
To optimise the model efficiently, we employ Group Relative Policy Optimisation (GRPO) \citep{shao2024deepseekmathpushinglimitsmathematical} as an alternative to standard Proximal Policy Optimisation (PPO) \citep{schulman2017proximalpolicyoptimizationalgorithms}. GRPO eliminates the need for a value model (critic) by estimating advantages from group scores, thereby reducing computational costs associated with a critic model.

\paragraph{Hyper-parameter Selection}
For the purpose of validating our proposed approach, we adopt a default hyper-parameter settings set by vLLM \citep{kwon2023efficientmemorymanagementlarge}. Our primary objective is to demonstrate the viability of applying our framework onto LLMs, rather than to achieve optimal performance. Although more extensive hyper-parameter tuning may yield improved results, we fix the policy's (actor) temperature, nucleus sampling (top-p) and gradient clipping at 1, as our aim is to demonstrate the feasibility of low-cost neurosymbolic training pipelines and motivate further exploration in this direction.

\paragraph{Training Setting}
We fine-tune our pre-trained model using a constant learning rate of $1 \times 10^{-6}$ and a KL divergence penalty weighted at 0.001. Training is performed in bfloat16 precision on 2 NVIDIA A100 80GB GPUs with a batch size of 256 and both the input and output sequence length are capped at 2048. In our GRPO setup, we sample 16 rollouts per prompt and estimate the advantage by using the group's reward mean and standard deviation as a baseline. These settings reflect our available hardware resources.


\subsection{Evaluation and Performance Metrics}
\label{subsec:evaluation and performance metrics}
Our evaluation protocol quantifies the performance of SMT constraint generation through automated, reproducible measures:

\paragraph{Automated Verification}
Let \(\varphi^{\star}\) be the ground‑truth constraint and \(\hat{\varphi}\) the constraint generated by the model.  
We deem them correct when they are logically equivalent:
\begin{equation}
\label{eq:constraint checking}
\varphi^{\star} \;\Leftrightarrow\; \hat{\varphi}
\;\;\Longleftrightarrow\;\;
\Bigl[(\varphi^{\star} \land \lnot \hat{\varphi}) \text{ is UNSAT} \;\wedge\; (\hat{\varphi} \land \lnot \varphi^{\star}) \text{ is UNSAT}\Bigr]
\end{equation}

This bidirectional UNSAT test guarantees that no interpretation satisfies one constraint while violating the other, thereby certifying semantic equivalence.  
The resulting boolean signal is consumed by the reward function described in \autoref{sec:rl-enhanced model}.

\paragraph{Accuracy Metric} 
We define accuracy as the proportion of tasks for which the generated SMT constraint is both syntactically valid and semantically correct (i.e., successfully verified by Z3). By measuring solver-verified equivalence to the true worst-case execution condition, our metric eschews surrogate objectives such as BLEU or perplexity, which offer no guarantee of formal correctness.
\section{Experiments}
\label{sec:experiments}

We evaluate \textbf{WARP}, our neurosymbolic approach for \emph{incremental worst-case constraint reasoning}, across a wide range of popular large language models to assess its general applicability. 
In addition, we present \texttt{WARP-1.0-3B}, a fine-tuned model trained with the WARP approach, which achieves the best performance in its size range and even surpasses several larger open-source and proprietary baselines. 
Together, these evaluations demonstrate both the broad effectiveness of WARP when applied to existing models and its ability to lift performance across scales, enabling models of different sizes to reach levels typically reserved for much larger ones.
The evaluation of WARP is organised around three research questions:

\begin{description}
  \item[RQ1.] How does incremental worst-case constraint reasoning enable LLMs to outperform vanilla prompting, particularly when extrapolating from small input sizes to larger ones? 
  \item[RQ2.] To what extent does WARP’s performance improvement reflect genuine symbolic reasoning ability rather than domain-specific adaptation to SMT-LIB syntax?
  \item[RQ3.] How robust is incremental worst-case constraint reasoning across different model scales, varying extrapolation regimes, and program families representative of real software workloads, and what practical trade-offs emerge in terms of accuracy, scalability, and computational cost?
\end{description}

\subsection{Evaluation Setup}
\label{subsec:experiment setup}


\paragraph{Evaluation Prompt}
We assess each model’s ability to produce complete, structured SMT constraints by using the uniform instruction set (see \autoref{tab:warp-evaluation-prompts}). This instruction set directs the model to present its full reasoning between \texttt{<think>…</think>} tags and to return an exact, canonical SMT assertion between \texttt{<answer>…</answer>} tags, without any additional commentary or truncation. All evaluations are performed zero-shot, with no imposed limit on output length.

\paragraph{Baselines}
We benchmark \texttt{WARP‑1.0-3B} against nine open-source checkpoints (3B–32B) and several contemporaneous OpenAI models, selected to span a representative range of model scales within our computational budget.


\lstset{
  basicstyle=\ttfamily\footnotesize, 
  breaklines=true,
  breakatwhitespace=true,
  postbreak=\mbox{},
  breakindent=0pt,
  frame=lines,
  escapeinside={(*@}{@*)},
}

\begin{lstlisting}[float,caption={Instruction set for evaluating WARP-1.0. (\textcolor{red}{\textbf{[\texttt{QUESTION}]}}) is a single prompt from the benchmark (see \autoref{fig:warp-benchmark instance}).},label={tab:warp-evaluation-prompts}]
You are a helpful assistant.
User: All per-variable constraints must be combined using a top-level (assert (and ...)) clause.
The output must be in exact, canonical SMT-LIB format without extra commentary in the constraint string.
Show your work in <think> </think> tags. And return the final SMT-LIB constraint string in <answer> </answer> tags.
For example: <answer>(assert (and  ( >=  in0 97)  ( <=  in0 122)))</answer>.
(*@\textcolor{red}{\textbf{[QUESTION]}}@*)
\end{lstlisting}

\begin{table}
\caption{Symbolic constraint generalisation accuracy (\%) across models evaluated under the WARP incremental worst-case constraint reasoning approach. 
In general, larger and more sophisticated LLMs achieve higher accuracy on worst-case constraint inference, and all models perform substantially better than the vanilla prompting results reported earlier (\autoref{subsec:RQ1}). 
The standout result is \texttt{WARP-1.0-3B} (marked with $\ast$), fine-tuned from the lowest-performing baseline, \texttt{Qwen2.5-3B-Instruct}, which delivers more than a fourfold improvement over its base model and performs on par with, or better than,  much larger open-source and proprietary baselines.}
\label{tab:model_results}
\centering
\footnotesize
\begin{tabularx}{\textwidth}{>{\bfseries}l 
    >{\raggedleft\arraybackslash}X 
    >{\raggedleft\arraybackslash}X 
    >{\raggedleft\arraybackslash}X 
    >{\raggedleft\arraybackslash}X}
\toprule
Model & Trial 1 (\%) & Trial 2 (\%) & Trial 3 (\%) & Average (\%) \\
\midrule
Qwen2.5-3B-Instruct       & 11.18 & 9.99  & 8.20  & 9.79 (±1.49) \\
DeepSeek-R1-Distill-Qwen-7B    & 18.63 & 16.39 & 16.24 & 17.09 (±1.30) \\
Qwen2.5-7B-Instruct            & 21.46 & 25.04 & 22.95 & 23.15 (±1.79) \\
Falcon3-10B-Instruct           & 28.02 & 29.81 & 29.36 & 29.06 (±0.90) \\
GPT-4o-2024-11-20              & 37.26 & 36.96 & 33.08 & 35.77 (±2.33) \\
DeepSeek-R1-Distill-Qwen-14B   & 36.51 & 36.36 & 35.32 & 36.07 (±0.66) \\
WARP-1.0-3B $\ast$             & 41.43 & 41.58 & 40.24 & {41.08 (±0.72)} \\
GPT-4.1-2025-04-14             & 44.11 & 45.31 & 42.47 & 43.96 (±1.44) \\
Qwen2.5-14B-Instruct           & 44.86 & 46.20 & 48.73 & 46.60 (±1.94) \\
DeepSeek-R1-Distill-Qwen-32B   & 48.88 & 50.52 & 52.91 & 50.77 (±2.02) \\
GPT-4.1-mini-2025-04-14        & 57.82 & 59.61 & 55.89 & 57.77 (±1.86) \\
Qwen2.5-32B-Instruct           & 70.19 & 70.64 & 68.70 & 69.85 (±1.05) \\
o4-mini-2025-04-16             & 86.89 & 87.78 & 90.61 & 88.43 (±1.52) \\
GPT-5-2025-08-07               & 95.83 & 95.08 & 95.50 & 95.47 (±0.38) \\
\bottomrule
\end{tabularx}
\end{table}

\subsection{Overall Results}

Our overall results in \autoref{tab:model_results} first demonstrate that incremental worst-case constraint reasoning consistently equips LLMs to outperform vanilla prompting. 
While GPT-5 represents the strongest proprietary baseline, its performance under vanilla prompting falls short of all models evaluated with WARP when equipped with the incremental reasoning strategy. 
Specifically, GPT-5 achieves only 13\% accuracy when prompted directly, but reaches 95.47\% under WARP, underscoring that incremental worst-case constraint reasoning is essential for enabling reliable generalisation.

The fine-tuned model \texttt{WARP-1.0-3B} achieves an average zero-shot generalisation accuracy of \(41.08\%\) on our worst-case symbolic-constraint benchmark, evaluated over three independent trials. This performance represents a more than fourfold improvement over the instruct-tuned \texttt{Qwen2.5-3B-Instruct} baseline (\(9.79\%\)) and nearly doubles the accuracy of DeepSeek-7B (\(17.09\%\)). Furthermore, \texttt{WARP-1.0-3B} surpasses mid-scale models such as \texttt{Falcon3-10B} (\(29.06\%\)) and DeepSeek-14B (\(36.07\%\)), while maintaining a trial-to-trial variance of less than \(1\) percentage point, indicating strong stability and consistent solver alignment. \autoref{tab:model_results} presents the zero-shot generalisation results for all models, including DeepSeek‐R1‐Distill‐Qwen variants with 7B, 14B, and 32B parameters (hereafter DeepSeek-7B, DeepSeek-14B, and DeepSeek-32B, respectively).

The recently released \texttt{GPT-5} achieves an average of \(95.47\%\) on our benchmark when combined with the incremental worst-case constraint reasoning strategy, whereas applied directly it performs poorly (see \autoref{tab: vanilla prompting results}). In parallel, our solver-aligned reinforcement learning framework (RLSGF) produces large gains at the small scale: \texttt{WARP-1.0-3B} improves from \(9.79\%\) (instruct variant of base model) to \(41.08\%\). These results establish both the upper bound provided by frontier models and the relative improvement enabled by RLSGF at compact scales.

\autoref{fig:model-accuracy} depicts each model’s average accuracy plotted against parameter count on a log scale. Compact baselines such as \texttt{Qwen2.5-3B} (\(9.79\%\)) and \texttt{DeepSeek-7B} (\(17.09\%\)) lie at the lower end, while \texttt{Qwen2.5-7B} (\(23.15\%\)), \texttt{Falcon3-10B} (\(29.06\%\)), and mid-sized open-source models like \texttt{DeepSeek-14B} (\(36.07\%\)) and \texttt{Qwen2.5-14B} (\(46.60\%\)) occupy the middle. Among the largest open models, \texttt{DeepSeek-32B} scores \(50.77\%\) and \texttt{Qwen2.5-32B} scores \(69.85\%\), alongside proprietary model GPT-4o (\(35.77\%\)) and its successor GPT-4.1 (\(43.96\%\)), whose sizes are not publicly disclosed but are believed to be comparable. Notably, \texttt{WARP-1.0-3B} outperforms its parameter-size peers and matches or surpasses several \(14\)B models. \texttt{WARP-1.0}’s deviation from parameter-only scaling highlights the significant impact of RLSGF, demonstrating that integrating formal SMT validation into the fine-tuning loop enables compact LLMs to synthesise semantically precise worst-case path constraints, achieving fidelity comparable to much larger models.

\begin{figure}[htp]
  \centering
  \begin{tikzpicture}
    \begin{axis}[
        xmode=log,
        log basis x=10,
        width=0.8\textwidth,
        height=0.5\textwidth,
        xlabel={Model Size (Billion Parameters)},
        ylabel={Average Accuracy (\%)},
        grid=major,
        xmin=1, xmax=200,
        ymin=0, ymax=100,
        xtick={1, 3, 10, 30, 100, 175},
        xticklabels={1, 3, 10, 30, 100, 175},
        ytick={0,10,20,30,40,50,60,70,80,90,100},
        legend pos=south east,
        legend cell align=left,
        scatter/use mapped color={draw=Plum, fill=Plum},
        clip=false
      ]

      \addplot+[
        only marks,
        mark=*,
        black,
        visualization depends on={value \thisrow{label} \as \Label},
        nodes near coords*={\Label},
        every node near coord/.append style={
          font=\scriptsize,
          anchor=west,
          xshift=3pt,
          inner sep=0pt
        },
      ]
      table[x=x,y=y,col sep=space] {models.dat};

      \addplot[thin, black] coordinates {(1,35.77) (200,35.77)};
      \node[anchor=west, font=\scriptsize] at (axis cs:200,35.77) {GPT-4o$^{\dagger}$};

      \addplot[thin, black] coordinates {(1,43.96) (200,43.96)};
      \node[anchor=west, font=\scriptsize] at (axis cs:200,43.96) {GPT-4.1$^{\dagger}$};

      \addplot[thin, black] coordinates {(1,57.77) (200,57.77)};
      \node[anchor=west, font=\scriptsize] at (axis cs:200,57.77) {GPT-4.1-mini$^{\dagger}$};

      \addplot[thin, black] coordinates {(1,88.43) (200,88.43)};
      \node[anchor=west, font=\scriptsize] at (axis cs:200,88.43) {o4-mini$^{\dagger}$};

      \addplot[thin, black] coordinates {(1,95.47) (200,95.47)};
      \node[anchor=west, font=\scriptsize] at (axis cs:200,95.47) {GPT-5$^{\dagger}$};

    \end{axis}
  \end{tikzpicture}

  \caption{Model size vs.\ symbolic constraint accuracy. Points show models with publicly known parameter counts; dashed lines show benchmarks for models with undisclosed sizes.~{\footnotesize$^{\dagger}$Parameter sizes for these models are proprietary / not publicly disclosed. Horizontal lines indicate reported average accuracies only.}}
  \label{fig:model-accuracy}
  
\end{figure}

\begin{table*}
\footnotesize
\centering
\caption{Error-type breakdown (aggregated over three trials). \texttt{Correct} denotes solver-equivalent outputs; \texttt{Syntax} are invalid SMT constraints; \texttt{Semantics} are syntactically valid but logically incorrect constraints; \texttt{Formatting} indicates structural issues (e.g., tag mismatches).}
\label{tab:error_breakdown}
\begin{tabularx}{\textwidth}{l 
    >{\raggedleft\arraybackslash}X 
    >{\raggedleft\arraybackslash}X 
    >{\raggedleft\arraybackslash}X 
    >{\raggedleft\arraybackslash}X}
\toprule
Model & Correct & Syntax & Semantics & Formatting \\
\midrule
Qwen2.5-3B      & 197  & 1626 & 135 & 55  \\
Qwen2.5-7B      & 466  & 1388 & 151 & 8   \\
Qwen2.5-14B     & 938  & 942  & 126 & 7   \\
Qwen2.5-32B     & 1406 & 468  & 139 & 0   \\
DeepSeek-7B     & 344  & 110  & 213 & 356 \\
DeepSeek-14B    & 726  & 1044 & 110 & 133 \\
DeepSeek-32B    & 1022 & 827  & 134 & 30  \\
Falcon-10B      & 585  & 1331 & 88  & 9   \\
WARP-1.0-3B    & 827  & 275  & 497 & 414 \\
GPT-4.1         & 885  & 455  & 672 & 1   \\
GPT-4.1-mini    & 1163 & 305  & 543 & 2   \\
GPT-4o          & 720  & 464  & 675 & 154 \\
o4-mini         & 1780 & 21   & 196 & 16  \\
GPT-5           & 1919 & 15   & 79  & 0   \\
\bottomrule
\end{tabularx}
\end{table*}

\begin{tcolorbox}[colback=gray!5,colframe=gray!40!black,title=Answer to RQ1]
The results confirm that incremental worst-case constraint reasoning provides a consistent advantage over vanilla prompting for LLMs. 
After fine-tuning, \texttt{WARP-1.0-3B} achieves more than four times the accuracy of its base model and performs competitively with models an order of magnitude larger. 
These findings show that incremental reasoning is key to achieving reliable generalisation across model scales.
\end{tcolorbox}

\subsection{Disentangling Symbolic Reasoning from Syntax Adaptation}
\label{subsec:RQ2}

In \autoref{tab:model_results}, WARP's solver-aligned framework improves accuracy in \texttt{WARP-1.0-3B} over its base \texttt{Qwen2.5-3B-Instruct} model from 9\% to 41\%. Such gains are substantial, but raw accuracy could potentially mask whether the model has genuinely learned symbolic reasoning or is merely adapting to the SMT format. Disentangling these factors is essential for understanding the true source of improvement and the extent to which WARP advances beyond surface-level pattern matching.

To make this distinction explicit, we aggregate outcomes into four categories: \emph{Correct}, \emph{Semantics}, \emph{Syntax}, and \emph{Formatting}. A prediction is labelled \emph{Correct} if it is solver-equivalent to the ground truth; \emph{Semantics} if it is syntactically valid but fails equivalence; \emph{Syntax} if it cannot be parsed by Z3; and \emph{Formatting} if the output fails to appear within the required \texttt{<answer>} block. These categories follow a natural dependency: semantic checks are only possible for syntactically valid outputs, and syntax can only be assessed if the format permits extraction. In practice, this means that some errors may conceal deeper correctness, while others expose fundamental weaknesses in instruction-following.

\subsubsection{Aggregated Results}
As shown in \autoref{tab:error_breakdown}, \texttt{WARP-1.0-3B} achieves a dramatic reduction in syntax errors compared to its instruct variant base model (\texttt{Qwen2.5-3B-Instruct}) (from 83\% to 14\%), but also suffers a high rate of formatting errors (21\%). Because the categories are dependent, many outputs never reach semantic evaluation, making raw counts an incomplete picture of reasoning ability. In particular, it remains unclear whether \texttt{WARP-1.0-3B}’s apparent semantic improvements stem from genuine generalisation or are masked by formatting fragility. To address this, we turn to targeted case studies that allow us to study individual predictions.

\subsubsection{Case Studies}

To better interpret the error categories, we study representative cases where automated evaluation under- or over-estimates reasoning ability. We analyse the LLM reasoning rationale, which refers to the intermediate explanation generated before the final SMT constraint, typically a step-by-step account of how patterns from smaller input sizes are identified and extrapolated to larger $n$.

\paragraph{Case Study A: Improved symbolic reasoning in WARP-1.0-3B}
We first examine a case where the base model produces a syntactically valid but \emph{semantically incorrect} constraint for $N=28$, while WARP generates a solver-equivalent formula under the same prompt (see \autoref{tab:caseA}).
The base model’s response misidentifies the governing rule: it describes the sequence as “monotonically increasing” and encodes only a single chain of $\ge$ relations. This both reverses the intended comparison direction and omits the strict $<$ operator that enforces a strictly decreasing chain. Although syntactically valid, the output is semantically non-equivalent. In contrast, \texttt{WARP-1.0-3B} extrapolates the correct adjacency schema—“for each $i>0$: $in_i \le in_{i-1}$ \emph{and} $in_i < in_{i-1}$”—across all indices at $N=28$, preserving both weak and strict constraints and verifying as solver-equivalent. The contrast indicates uplift in symbolic reasoning: WARP induces and propagates the correct logical pattern, while the baseline over-fits surface cues (constraint counts, “increasing” phrasing) and generalises an incorrect schema.

\lstset{
  basicstyle=\ttfamily\footnotesize, 
  breaklines=true,
  breakatwhitespace=true,
  postbreak=\mbox{},
  breakindent=0pt,
  frame=lines,
  escapeinside={(*@}{@*)},
}

\begin{lstlisting}[caption={Case study A (ID 236, small): for $N=28$, the baseline \texttt{Qwen2.5-3B-Instruct} produces a \emph{semantically wrong} constraint by enforcing a monotonically \textbf{\color{red}increasing} sequence (e.g., $in_1 \geq in_0, in_2 \geq in_1, \dots$), which reverses the intended logic. 
In contrast, \texttt{WARP-1.0-3B} correctly generates solver-equivalent constraints that enforce a monotonically \textbf{\color{blue}decreasing} pattern with the proper combination of weak and strict inequalities.},label={tab:caseA}]
(*@\textbf{Base Model Response:}@*)
To generate the constraint for N=28, we follow the same pattern as the previous examples. We have 28 invariants (constraints) to ensure that the input elements are monotonically increasing. This means we have:- 1 constraint for in1 being greater than or equal to in0.- Then 27 additional constraints ensuring in2, in3, ..., in28 are greater than their preceding elements. (*@$\langle$@*)truncated model output to save paper space(*@$\rangle$@*) This corresponds to a single (and) clause with 27 inequalities, detailing each successive element being greater than its previous one: 
(*@{\color{red}<answer>(assert (and (>= in1 in0) (>= in2 in1) (>= in3 in2) (>= in4 in3) (>= in5 in4) (>= in6 in5) (>= in7 in6) (>= in8 in7) (>= in9 in8) (>= in10 in9) (>= in11 in10) (>= in12 in11) (>= in13 in12) (>= in14 in13) (>= in15 in14) (>= in16 in15) (>= in17 in16) (>= in18 in17) (>= in19 in18) (>= in20 in19) (>= in21 in20) (>= in22 in21) (>= in23 in22) (>= in24 in23) (>= in25 in24) (>= in26 in25) (>= in27 in26)))</answer>}@*)

(*@\textbf{WARP-1.0-3B Response:}@*)
As the input size N increases, the constraints are built based on the previous constraints and a pattern emerges for each number. For instance, for N=4, we already have the set of constraints given. For N=8, we have an additional layer of brackets, and for N=10, we would need to generate an expanded set of constraints following the pattern seen in the smaller inputs. For N=28, we have an input that is double the N of N=27, and it should maintain the pattern observed in the previous inputs. The code for generating N=28 output constraints is almost the same as N=27, with minor adjustments for input formatting. As a whole, this is a simple pattern generating program. Here is an example of the constraints for N=28:
(*@{\color{blue}
<answer>(assert (and  ( <=  in1 in0)  ( <  in1 in0)  ( <=  in2 in1)  ( <  in2 in1)  ( <=  in3 in2)  ( <  in3 in2)  ( <=  in4 in3)  ( <  in4 in3)  ( <=  in5 in4)  ( <  in5 in4)  ( <=  in6 in5)  ( <  in6 in5)  ( <=  in7 in6)  ( <  in7 in6)  ( <=  in8 in7)  ( <  in8 in7)  ( <=  in9 in8)  ( <  in9 in8)  ( <=  in10 in9)  ( <  in10 in9)  ( <=  in11 in10)  ( <  in11 in10)  ( <=  in12 in11)  ( <  in12 in11)  ( <=  in13 in12)  ( <  in13 in12)  ( <=  in14 in13)  ( <  in14 in13)  ( <=  in15 in14)  ( <  in15 in14)  ( <=  in16 in15)  ( <  in16 in15)  ( <=  in17 in16)  ( <  in17 in16)  ( <=  in18 in17)  ( <  in18 in17)  ( <=  in19 in18)  ( <  in19 in18)  ( <=  in20 in19)  ( <  in20 in19)  ( <=  in21 in20)  ( <  in21 in20)  ( <=  in22 in21)  ( <  in22 in21)  ( <=  in23 in22)  ( <  in23 in22)  ( <=  in24 in23)  ( <  in24 in23)  ( <=  in25 in24)  ( <  in25 in24)  ( <=  in26 in25)  ( <  in26 in25)  ( <=  in27 in26)  ( <  in27 in26)))</answer>}@*)
\end{lstlisting}
%
%
%
\begin{lstlisting}[caption={Case study B (ID 621, medium): WARP-1.0-3B response with formatting error. The output lacks the required <answer> tokens, but the extracted constraints are solver-equivalent.},label={tab:caseB}]
(*@\textbf{WARP-1.0-3B output:}@*)
To generate the constraint for N=22, we can extend the logic applied in the previous examples. We notice a pattern where the input values are permuted and checked against a set of constraints based on the current input and the previous ones. This pattern can be applied to larger inputs as well.  Here is the derived constraint for N=22:
Answer: (assert (and  (not ( = in0 65))  ( =  in0 66)  (not ( = in1 69))  ( =  in1 70)  
(not ( = in2 73))  (not ( = in2 74))  (not ( = in3 65))  ( =  in3 66)  
(not ( = in4 69))  ( =  in4 70)  (not ( = in5 73))  (not ( = in5 74))  
(not ( = in6 65))  ( =  in6 66)  (not ( = in7 69))  ( =  in7 70)  
(not ( = in8 73))  (not ( = in8 74))  (not ( = in9 65))  ( =  in9 66)  
(not ( = in10 69))  ( =  in10 70)  (not ( = in11 73))  (not ( = in11 74))  
(not ( = in12 65))  ( =  in12 66)  (not ( = in13 69))  ( =  in13 70)  
(not ( = in14 73))  (not ( = in14 74))  (not ( = in15 65))  ( =  in15 66)  
(not ( = in16 69))  ( =  in16 70)  (not ( = in17 73))  (not ( = in17 74))  
(not ( = in18 65))  ( =  in18 66)  (not ( = in19 69))  ( =  in19 70)  
(not ( = in20 73))  (not ( = in20 74))  (not ( = in21 65))  ( =  in21 66)))

(*@\textbf{Evaluation outcome:} No <answer> tokens detected;\\
\textbf{Solver check:} equivalent.@*)
\end{lstlisting}

\paragraph{Case Study B: Formatting repair in WARP-1.0-3B}
Unlike the previous case study, which highlights symbolic reasoning, this example concerns adherence to output format rules. As described in \autoref{subsec:experiment setup}, each response is expected to enclose the synthesised constraints within \texttt{<answer>} and \texttt{</answer>} tokens. When the model omits these delimiters, produces multiple spans, or places extraneous text outside them, evaluation cannot parse the result—even if the constraints themselves are valid.
From \autoref{tab:error_breakdown}, \texttt{WARP-1.0-3B} shows the highest rate of such formatting errors. This stems from its base architecture (\texttt{Qwen2.5-3B}), which was trained without supervised instruction tuning. While reinforcement learning enabled effective symbolic generalisation, it made the model less reliable at following strict protocol instructions. Other instruction-tuned baselines adhere more consistently to the template, even if their SMT constraints are semantically wrong. This case study (see \autoref{tab:caseB}), therefore shows that \texttt{WARP-1.0-3B} can generate correct constraints but often fails to package them in the required format.

In conclusion, \texttt{WARP-1.0-3B} shows instances of genuine logical generalisation beyond surface patterns, the dependence on strict formatting and extraction rules means that some outputs may be categorised as errors despite containing solver-equivalent constraints. These results suggest that WARP advances symbolic reasoning ability, but the evaluation remains sensitive to categorisation and protocol adherence.

\begin{tcolorbox}[colback=gray!5,colframe=gray!40!black,title=Answer to RQ2]
WARP’s performance gains are not explained by simple adaptation to SMT syntax alone. 
While fine-tuning substantially reduces syntax errors compared to the base model, \texttt{WARP-1.0-3B} also achieves a much higher proportion of solver-equivalent outputs, demonstrating improved semantic correctness. 
Case studies further confirm that WARP corrects logically invalid constraints produced by the baseline, showing that reinforcement learning with solver-guided feedback enables genuine symbolic reasoning rather than surface-level pattern mimicry.
\end{tcolorbox}

\subsection{Robustness Across Scales, Regimes and Program Families} 
\label{subsec:RQ3}

We evaluate the effectiveness of incremental worst-case–constraint reasoning as: (1) \emph{model scale} varies, (2) under \emph{extrapolation regimes} induced by the benchmark’s jump-size tiers, and (3) across \emph{program families} representative of real software—while remaining explicit about practical limits in accuracy, scalability, and computational cost.

\subsubsection{Across Model Scale}

Accuracy increases with parameter count, and the WARP framework disproportionately lifts compact models beyond parameter-only expectations (see \autoref{tab:model_results}). Failure modes also evolve with scale. Using the four labels from \autoref{subsec:RQ2} (Correct, Semantics, Syntax, Formatting), small models are dominated by syntax errors; mid-scale models reduce syntax but still incur non-semantic failures; in the most capable models, residual errors are predominantly semantic. For brevity we reference the aggregate breakdown in \autoref{tab:error_breakdown} and focus below on how these failures play out across extrapolation tiers and program families.

\subsubsection{Across Extrapolation Regimes}

\begin{table}[htp]
\footnotesize
\centering
\caption{Tier-wise correct counts (small/medium/large) and totals for representative models (aggregated over three trials).}
\label{tab:tier-by-model}
\begin{tabularx}{\textwidth}{l 
    >{\raggedleft\arraybackslash}X 
    >{\raggedleft\arraybackslash}X 
    >{\raggedleft\arraybackslash}X 
    >{\raggedleft\arraybackslash}X}
\toprule
                       Model &  Small &  Medium &  Large &  TotalCorrect \\
\midrule
                 DeepSeek-7B &    190 &     154 &      0 &           344 \\
                DeepSeek-14B &    373 &     349 &      4 &           726 \\
                DeepSeek-32B &    517 &     497 &      8 &          1022 \\
                 WARP-1.0-3B &    419 &     401 &      7 &           827 \\
                  Qwen2.5-3B &    119 &      78 &      0 &           197 \\
                  Qwen2.5-7B &    257 &     206 &      3 &           466 \\
                 Qwen2.5-14B &    467 &     465 &      6 &           938 \\
                 Qwen2.5-32B &    683 &     712 &     11 &          1406 \\
                 Falcon3-10B &    308 &     275 &      2 &           585 \\
                       GPT-5 &    961 &     943 &     15 &          1919 \\
                GPT-4.1-mini &    613 &     542 &      8 &          1163 \\
                     GPT-4.1 &    457 &     423 &      5 &           885 \\
                     o4-mini &    890 &     876 &     14 &          1780 \\
                      GPT-4o &    422 &     294 &      4 &           720 \\
\bottomrule
\end{tabularx}
\end{table}

Although training data are not labelled by difficulty, the evaluation partitions instances by jump size (small/medium/large), providing natural regimes to probe generalisation as extrapolation increases. Empirically, accuracy degrades gently from small to medium, then drops on the large tier for most models; only the most capable systems sustain high accuracy. \autoref{tab:tier-by-model} summarises tier-wise correct counts. Read with \autoref{tab:error_breakdown}, this indicates regime-specific brittleness consistent with longer SMT outputs and compounding errors at large jumps.

\subsubsection{Across Program Families}
Performance differs by program family, reflecting real-world variability. The hardest cases in our benchmark are Dijkstra (3.57\% micro-accuracy overall) and BinarySearch (8.75\% overall; local models 0.60\%). Strongly regular patterns such as TowerOfHanoi and MazeSolver are broadly solved (88.99\% and 85.92\% overall). Best single-model accuracy ranges from 50.00\% on Dijkstra and 66.67\% on BinarySearch to 100\% on several regular families, indicating genuine cross-family variability even as overall accuracy rises.

\begin{tcolorbox}[colback=gray!5,colframe=gray!40!black,title=Answer to RQ3]
Incremental worst-case constraint reasoning is robust across scales, extrapolation regimes, and
program families. The WARP approach lifts smaller models disproportionately. Accuracy does degrade smoothly on medium jumps and more sharply on large ones. Performance varies by program type, with irregular graph workloads proving hardest while regular patterns are broadly solved. These findings highlight genuine cross-family generalisation alongside practical trade-offs in token, verification, and compute limits.

\end{tcolorbox}

\subsection{Threats to Validity}
\label{subsec:threats}

\paragraph{Internal Validity.}  
The correctness of our evaluation depends on automated equivalence checking with the Z3 solver. 
Formatting or parsing errors may prevent valid constraints from being assessed, potentially underestimating reasoning ability. 
In addition, the use of a uniform prompt template may advantage instruction-tuned models, introducing bias into relative performance comparisons.

\paragraph{Construct Validity.} We measure accuracy as the proportion of outputs that are both syntactically valid and solver-equivalent to the ground truth. This binary signal, while being rigours, may mask partial progress, such as outputs that capture much of the intended structure but deviate slightly. We mitigate this by including case studies that expose under- and over-estimations of reasoning ability, but finer-grained metrics could provide additional insight.

\paragraph{External Validity.}  
Our benchmark consists of 20 Java programs and input sizes up to $n=30$. 
Although these cover a variety of algorithms, they cannot represent the full spectrum of real-world software or larger-scale symbolic reasoning tasks, particularly those involving deeper recursion or more complex data structures. 

\paragraph{Token Length Limits.}  
All models are constrained by maximum context and generation lengths. 
This restricts both the size of prompts and the completeness of outputs, particularly for larger $n$, where constraints may exceed sequence limits. 
This highlights an inherent scalability challenge for autoregressive LLMs when applied to symbolic constraint reasoning.
\\\\
These threats do not undermine the main contributions of this work: formalising incremental worst-case constraint reasoning, introducing the WARP approach with solver-guided feedback, and releasing both a benchmark and a proof-of-concept model. 
Rather, they define the boundaries of the current evaluation. 
Future work will extend WARP fine-tuning to larger open-source LLMs, explore alternative constraint generation and evaluation strategies, and investigate broader program domains, establishing best practices for neuro-symbolic program constraint reasoning.

\section{Conclusion} 
\label{sec:conclusion}
In this work, we report on our neurosymbolic approach for computing and generalising worst-case symbolic constraints for Java programs. As part of this work, we developed \texttt{WARP-1.0-3B}, which achieves performance that exceeds previous generation state-of-the-art models like \texttt{GPT-4o}, and lays the foundation for a series of future research opportunities.
While the latest OpenAI thinking-series models achieve remarkable results on our benchmark, their success highlights the broader momentum behind scaling test-time reasoning. Our findings complement this trend, showing that a reinforcement learning approach such as RLSGF can equip compact models like \texttt{WARP-1.0-3B} with strong symbolic generalisation offering a lightweight and scalable path toward more capable neurosymbolic systems.

\section*{Data Availability}
The data that support the findings of this study, including all our results and models, are openly available in our supplemental material:

\url{https://huggingface.co/datasets/dannkoh/warp-training} 

\url{https://huggingface.co/datasets/dannkoh/WARP-benchmark}

\url{https://github.com/dannkoh/warp-veRL}

\url{https://github.com/dannkoh/WARP-evaluation}

\bibliographystyle{ACM-Reference-Format}
\bibliography{refs}




\end{document}